\documentstyle[12pt]{article}
\begin{document}

\title{Test of the hierarchical theory for the FQHE}
\author{M. Kasner \\Physikalisch--Technische
Bundesanstalt
\\Bundesallee 100
\\D--38116 Braunschweig, Germany}
\date{}
\maketitle
\begin{abstract}
In the hierarchical theory of the fractional quantum Hall effect,
the low--energy behaviour of a daughter state in the next level of the
hierarchy is described by an interacting system of quasiparticles of the
parent state. Taking the filled lowest Landau level as the parent state,
we examine analytically the quantitative consequences
of this approach for electrons interacting via a pseudopotential interaction.
It is shown that the ground
state energy per particle in the daughter state at a filling
factor $2/3$ is exactly equal to
that of a system of quasiholes in the parent state with half filling,
precisely as predicted by
the hierarchical approach. This is achieved with
{\em only up to two--particle interactions} in the effective
Hamiltonian for the quasiholes. Their single particle energy and
two--particle interaction are derived.
The results are generalized to the other filling factors
attainable from the filled Landau level.
\end{abstract}

\newpage
{\em Introduction.} All attempts to understand the surprising transport
properties in the
fractional quantum Hall effect
(FQHE) \cite{TSG82} are based on investigations
of the energy spectrum of interacting electrons in a strong magnetic field.
While the low--energy behaviour at filling factors $\nu =1/q$ is quite well
understood after Laughlin's work \cite{Lau83b}, the theory for the
occurence of filling factors $p/q$ ($p \neq 1$) is still under debate. A
sound theory should not only
explain their occurence but also make
more quantitative predictions about the
energy gap, the plateauwidth and consequently about the stability in the
experiment; e.\ g.\ \nolinebreak , the plateau of the 1/3--state is
stronger than that of the 2/7--state, and the 4/11--plateau has not been
observed for
spinpolarized electrons up to now \cite{DSTPW93}. There are at
least two approaches \cite{Mac92}. \\
One approach, the hierarchical theory, was originally introduced by
Halperin and Haldane \cite{Halp83,Hald83}, see also \cite{Gir84a,MAD85,MM85}.
There, the occurence of
the filling factors $p/q$ ($p \neq 1$) is explained by a qualitative
analogy to a
$1/q$-ground state, the Laughlin state, in which the energetically
degenerated electrons in the lowest Landau level form a particularly
stable state under the influence of their interaction. Small deviations
from the filling factor $1/q$ of this "parent state" create
energetically degenerated
quasiparticles (quasiholes and
quasielectrons, respectively). Then, for a macroscopic number of these
quasiparticles, again a condensation phenomenon occurs under the
influence of the two--quasiparticle interaction,
and a new "daughter--FQHE--state" is formed.
This scheme can be applied iteratively, so that finally all states with an
odd--denominator filling factor
are predicted by the theory. \\
The second approach, the composite fermion theory by Jain \cite{Jai89a},
avoids a hierarchical construction. Instead, the trial wave
functions at $\nu =p/q$ are explicitly given by the electronic degrees of
freedom, although not in a very convenient form. Energy
expectation values of these trial
wave functions show quite good agreement with exact data,
particularly on the sphere \cite{DJ92b}. However, the reason for the success of
these wave functions still remains unclear.\\
Either approach has its merits and it seems to be worthwhile to determine the
range of applicability of these theories by more quantitative calculations.\\
In order to check the hierarchical theory quantitatively,
one has to show that the low--energy behaviour of
the fermions at the filling factor of the daughter state can be described by
the quasiparticles moving in the background at the parent state. The first
and simplest test
is the comparison of the ground state
energy. In order to do this,
the
interaction of the quasiparticles has to be determined. There
are up to now two attempts to carry out such a
programme. B\'eran
and Morf calculated the two--particle interaction of the
$1/3$--quasielectrons forming
the $2/5$--daughter--state in the hierarchical picture with anyonic
quasiparticle
wave functions, and they compared their ground state and gap energies
with those of the
original system \cite{BM91}. Endesfelder and
Terzidis extracted the interaction potential coefficients for the
$1/3$--quasiholes from exact diagonalizations of small systems and compared
then the bosonic quasihole spectrum with the fermionic
spectrum at the filling factor $2/7$ \cite{ET92}. However, they did not
study the ground state energy at $2/7$ for a macroscopic number of
quasiparticles. \\
In this paper, we want to investigate the validity of the hierarchical
theory by comparing ground state energies resulting from two different
calculations. In the first calculation, we take the filled Landau level
as the parent state. Exciting $N/2$ quasiholes in the background of this state
results in a daughter state at filling $2/3$, according to the hierarchical
theory; $N \rightarrow \infty $ is the number of electrons in the parent state.
In order to calculate the ground state energy in this approach, we determine
the single particle
energy and the interaction in the effective quasihole Hamiltonian from exact
spectra of systems with one and two quasiholes,
respectively.
Then, considering this effective Hamiltonian now for $N/2$ quasiholes,
its ground state energy can be exactly determined, and from this, the
ground state energy
per particle of the daughter state at filling factor $2/3$ is derived in the
thermodynamic limit. In the second calculation, we derive the
goundstate energy at
filling factor $2/3$ from the well known ground state energy at $1/3$ using
the particle--hole--transformation. \\
In the following, we will perform our calculations in the general case of
$N/p$ quasiholes in the parent state which leads to a filling
factor of $p/(p+1)$ of the daughter state
($p$ even; $p$ is not to be confused with
the numerator from the filling factor $p/q$). The appeal of our
approach lies in
the fact that we determine all energies analytically. \\
Next, after introducing our model, we determine the ingredients in the
effective Hamiltonian, show that the electronic
states can be described by the bosonic quasihole Hamiltonian,
and do the comparison. \\

{\em The model.}
We investigate electrons moving on a two--dimensional disk. The single
particle Hilbert
space is restricted to spinpolarized states
in the lowest Landau level by a strong magnetic field, and the
only quantum number of these energetically degenerated
states is the angular momentum $m$ ($m=0,1,\ldots $).
The Hamiltonian of a finite system of interacting electrons
is
\begin{equation}
H  =  \frac{1}{2} \sum_{m_{1},m_{2}, \atop m_{3},m_{4}=0}^{m_{{\rm max}}}
W_{m_{1} m_{2} m_{3} m_{4}} c^{\dagger}_{m_{1}} c^{\dagger}_{m_{2}}
c_{m_{3}} c_{m_{4}} .
\end{equation}
The filling factor is given by the relation $\nu =\frac{N-1}{m_{{\rm max}}}=
\frac{N-1}{N_{\phi}-1}$, where $N$ is the number of electrons, $m_{{\rm max}}$
 is the
maximum single particle angular momentum and $N_{\phi}$ is the number of flux
quanta through the disk and thus defines its area. $N_{\phi}$ is also
the number of single particle states, i.\ e.\ , the degree of degeneracy.
The matrix elements
$W_{m_{1}m_{2}m_{3}m_{4}}$ are fixed by the two--particle interaction
$V(|z-z'|)$ ($z=x-iy$) which is assumed to be isotropic and
translationally invariant. As already pointed out by Haldane,
any interaction in the lowest Landau level
can be characterized by pseudopotential coefficients $V_{k}$
\cite{Hald90}. $k \geq 0$ denotes the relative angular momentum, even for
bosons, odd for fermions. Reversely, we
can construct an arbitrary interaction in the Hilbert space of the lowest
Landau level by a certain choice of the $V_{k}$.
It is well--known that the
Laughlin state $\Psi (z_{1},\ldots,z_{N})= \prod _{k < l}^{N} (z_{k}-
z_{l})^{q} \exp(-\frac{1}{4}\sum_{i=1}^{N} |z_{i}|^{2})$ is the exact
non--degenerate ground state of energy zero at filling factor $\nu =1/q$
( $q$ odd for fermions and even for bosons) for the following choice of the
interaction: the $V_{k}$ are arbitrary for $0 \leq k \leq q-1$
and $V_{k}=0$
 for all odd (even) $k$ with $k \ge q$ \cite{TK85,PT85}. For
our explicit calculations, we will use below the special interaction
parametrized by an even number $p \geq 2$  such that
\begin{eqnarray}
V_{k} > 0  \hspace{0.7cm} (0 \leq k \leq p) ; \hspace{1.0cm}  V_{k}=0
\hspace{1.0cm}    (k > p )  .
\end{eqnarray}
\\

{\em Two particles.} The energy spectrum of two fermions or bosons
moving in the
lowest Landau level can be described by the pseudopotential coefficients
$V_{k}$. The problem separates into a free motion of the center
of mass and a relative motion in the one--particle
potential $V(|z_{-}|)$ \hspace{0.2cm} [$z_{+}=(z_{1}+z_{2})/2,
z_{-}=z_{1}-z_{2}$; the
constant kinetic energy is
subtracted]. The total angular momentum $M$ and the relative and
center of mass
angular momenta $l_{r}$ and $l_{s}$ are conserved quantities
($M=l_{r}+l_{s}$).
The eigenvalues are independent of $l_{s}$ and given by
the pseudopotential coefficients $V_{l_{r}}$ with
$0 \leq l_{r} \leq M$
 (fermions: $l_{r}=2j-1$ -- odd, $1 \leq j \leq [\frac{M+1}{2}]$; bosons:
$l_{r}=2j$ -- even, $0 \leq j \leq [\frac{M}{2}]$; and
$[x]$ is the greatest integer not greater than $x$). So far, the
considerations apply to an infinite system, $m_{{\rm max}}=\infty$. In a
finite system, the eigenvalues of
angular momentum blocks of $H$ with $M \leq
m_{{\rm max}}$ are not influenced by the finiteness of the system (both single
particle angular momenta are then smaller than $m_{{\rm max}}$),
cf.\  \cite{KA93b},
and these eigenvalues can be
extracted from a two--particle spectrum without any finite size correction. \\

{\em   A single quasihole at $\nu=1$.} In order to introduce the notation,
we start with the filled lowest
Landau level, i.\ e.\ , a stable Laughlin state with $q=1$, where $N$
electrons occupy
the single particle states with angular momenta $m=0,\ldots,N-1$, i.\ e.\ ,
$N_{\phi }=N$ and $\nu =1$.  The total angular momentum of this state is
$M_{N}=\frac{1}{2}N(N-1)$, its total energy is denoted by
$E(N,N,M_{N}) \equiv E_{g}(N,N)$. Here and in the following, we provide
total energies with the arguments $N, N_{\Phi }$, and angular momentum $M$,
while $E_{g}(N,N_{\phi })$ denotes the ground state energy of a system with
$N$ particles and $N_{\phi }$ flux quanta.
The energy
per particle $\varepsilon _{0}(\nu =1)$ for this interacting electronic
system in the thermodynamic limit is determined from the definition of the
matrix elements $W_{m_{1}m_{2}m_{3}m_{4}}$
\begin{eqnarray}
\varepsilon _{0}(1)  & = & \lim_{N\to\infty}\frac{E_{g}(N,N)}{N}
\nonumber \\
& = & \lim_{N\to\infty} \frac{1}{2N} \sum_{m_{1},m_{2}=0}^{N-1}
(W_{m_{1}m_{2}m_{2}m_{1}}-W_{m_{1}m_{2}m_{1}m_{2}})   =
2 \sum_{i=1}^{p/2} V_{2i-1} .
\end{eqnarray}
The finite size correction decays as $1/\sqrt{N}$
\cite{KA93b}.
The last sum is cut off due to our special choice of the interaction
(2).
Here only quasiholes, but no quasielectrons, can be created
without leaving the lowest Landau level. Next, we
create a one--quasihole state by increasing the size of the system by one
flux quantum, i.\ e.\ ,  the
degeneracy $N_{\phi }$ by one.
The difference between the energy of these eigenstates with various
angular momenta $M$ and the
energy of the filled Landau level $\varepsilon _{-}^{n}(N,M;\nu
=1) \equiv E(N,N+1,M)-E_{g}(N,N)$ is defined as the neutral quasihole
energy (see for the different definitions of quasiparticle
energies \cite{MH86,MG86b}).
Because one of the single particle states is now unoccupied,
there are $N$ different
one--quasihole states with non-vanishing energy,
having total angular momenta $M$ from
$M=M_{N}+1$ to $M_{N}+N(\equiv M_{N+1})$. In the thermodynamic
limit, these
neutral one--quasihole energies become independent of $M$
and hence the quasihole states become energetically degenerated
just as for free
electrons in the lowest Landau level. Performing the summations over
occupied states (cf.\ (3)), we find for the
neutral quasihole energy
\begin{equation}
\varepsilon _{-}^{n}(1)=\lim_{N\to\infty} \varepsilon _{-}^{n}(N,M;1)
=-\varepsilon_{0}(1).
\end{equation}
Usually, the degeneracy is one of the basic assumptions of the
hierarchical theory. Here, it follows straight forwardly. \\

{\em Two quasiholes at $\nu =1$.} We next create a second quasihole by an
additional increase of the system size by one flux quantum, i.\ e.\ , only $N$
out of the now $N_{\phi }=N+2$ states are occupied by electrons.
There are angular momentum blocks with $M=M_{N}+2$ to
$M_{N}+2N( \equiv M_{N+2} -1)$.
The $i^{th}$ energy eigenvalue in a block $M$ is denoted by
$E_{i}(N,N+2,M)$ with $1 \leq i \leq min([\frac{M_{N+2}-M+1}{2}],
N+1-[\frac{M_{N+2}-M+2}{2}])$.
The eigenvalues of this fermionic system can be determined simply using
the particle--hole--transformation \cite{FO88} by relating the spectrum
of $N$ electrons at degeneracy
$N_{\Phi }$ to that of $N_{\Phi }-N$ holes in the Landau level at the same
degeneracy.
Performing the unitary particle--hole--transformation $U_{ph}$
we get the hole Hamiltonian
\begin{eqnarray}
H'  =  U_{ph}^{\dagger}HU_{ph}
& = & \frac{1}{2} \sum_{m_{1},m_{2}, \atop
m_{3},m_{4}=0}^{m_{{\rm max}}} W_{m_{1} m_{2} m_{3} m_{4}}
c'^{\dagger}_{m_{1}} c'^{\dagger}_{m_{2}}
c'_{m_{3}} c'_{m_{4}} \nonumber \\ -
\sum_{m_{1},m_{2}=0}^{m_{{\rm max}}} (W_{m_{1} m_{2} m_{2} m_{1}} & - &
W_{m_{1} m_{2} m_{1} m_{2}})(c'^{\dagger}_{m_{1}}
c'_{m_{1}} - \frac{1}{2})
\end{eqnarray}
with hole creation operators
$c_{m}'^{\dagger}=U_{ph}^{\dagger}c_{m}
U_{ph}$. The eigenvalues of the Hamiltonian $H$ (1)
are identical with those of $H'$ (5). If we specifiy $N_{\phi }=N+2$,
the eigenvalues of the two--quasihole block of $H$ are connected via
(5) with a two--particle--spectrum, for which the terms on the r.\ h.\ s.\
are known. We find
\begin{eqnarray}
&  & E_{i}(N,N+2,M)  \nonumber \\
& = & V_{2i-1} -  B_{i}(2,N+2,M_{N+2}-M)   +
E_{g}(N+2,N+2)
\end{eqnarray}
for all $M$ with $M_{N+1} \leq M \leq M_{N+2}-1$. The background
contribution
$B_{i}(g,N,M)$ coming from the one--particle term in (5) describes the
interaction of a state of g
holes of angular momentum $M$ with the homogenous background enclosing
       $N$
flux quanta. For its thermodynamic limit, we get for a two--hole state
in leading order (again performing the summations)
\begin{equation}
\lim_{N \to \infty} B_{i}(2,N,M)=4 \varepsilon _{0}(1).
\end{equation}
Thus, starting from the largest $M$, i.\ e.\ , from $M=M_{N+2}-1$, at
which there is only one eigenvalue in the block, $i=1$, we find
succesively the
corresponding fermionic eigenvalues $E_{i}(N,N+2,M)$. \\
The behaviour of all terms in (6) with respect to large increasing $N$ is
known. The above analytical determination of the energies is a special
advantage of the
case $\nu =1$. For other filling factors these eigenvalues had to be
extracted from numerically calculated spectra showing an {\em a priori}
unknown $N$--dependence.
\\

{\em Mapping to bosons.}
We want to describe the
fermionic spectra effectively by quasiholes moving in the
background of the filled Landau level and interacting via a
two--particle interaction $\tilde{V}(|z-z'|)$ parametrized by
pseudopotential coefficients $\tilde{V}_{k}$ which are to be determined.
What kind of particles are these quasiholes? We treat them
as bosons. This is justified for two quasiholes because not only the total
dimension of the two--quasihole Hilbert space of bosons, but also the
number of states in the sub-Hilbert
spaces with a definite total angular momentum $M$, are equal to those of the
fermionic system, as we show now. There are $N$
one--quasihole--states, i.\ e.\ , there are
$ {N+1 \choose 2} $  bosonic two--quasihole--states,
while the total dimension of a fermionic Hilbert space of $N$
electrons with $N+2$ one--particle--states is $ { N+1 \choose 2} $
(the outermost one--particle--state must be occupied). Hence the total
dimension of the two--quasihole Hilbert
space is the same in the fermionic and bosonic description, respectively. This
justifies the attempt
to regard the quasiholes as bosons
\cite{KA93b,HXZ92}.
Next, even the dimension of
a block in the two--quasihole Hamiltonian with fermionic total angular
momentum $M$ is the same as the dimension of a block in the two--boson
Hamiltonian with total angular momentum $\tilde{M}$ to be determined.
The fermionic dimension
for angular momenta with
$M_{N+2}-N \leq M \leq M_{N+2}-1$ is $[\frac{M_{N+2}-M+1}{2}]$ (see above two
quasiholes). On the other hand, the dimension
of a Hilbert space of two bosons with total angular momentum $\tilde{M}$
is  $[\frac{(\tilde{M}+2)}{2}]$. Thus,
for $\tilde{M}=M_{N+2}-M-1$ the
dimensions coincide. Therefore, the block of the fermionic Hamiltonian
with $M=M_{N+2}-1$ is mapped to a
block of a bosonic Hamiltonian with $\tilde{M}=0$, $M_{N+2}-2$
to $\tilde{M}=1$, $M_{N+2}-3$ to
$\tilde{M}=2$ etc.\ . Thus, for each given angular momentum, the two--quasihole
sub--Hilbert space can be mapped onto a two--boson sub--Hilbert space. Our
considerations can be easliy generalized to the case of $N$ quasiholes. \\
It should be mentioned that the character of the
quasiholes is intimately connected with the way we create them. We created
quasiholes by increasing the system size. If we would
create
quasiholes by taking electrons out of the system keeping the area fixed, then
these
so-called gross quasiholes are nothing but the holes in the section above,
i.\ e.\ , are fermions, see also
\cite{Hald91,JC93}. \\
Let us now make the following Ansatz for an effective bosonic Hamiltonian
\begin{equation}
\tilde{H}= \tilde{E}(N) + \sum_{m=0} \tilde{\varepsilon} (N,m)
b^{\dagger}_{m}b_{m} + \frac{1}{2} \sum_{m_{1},m_{2}, \atop m_{3},m_{4}=0}
\tilde{W}_{m_{1}m_{2}m_{3}m_{4}} b^{\dagger}_{m_{1}} b^{\dagger}_{m_{2}}
b_{m_{3}} b_{m_{4}},
\end{equation}
where $b_{m}^{\dagger}$ creates a boson with angular momentum $m$ ($m \ge 0$)
in the lowest Landau level.
The unknown
matrix elements $\tilde{E}, \tilde{\varepsilon }, \tilde{W}$ can be
determined by the condition that the eigenvalues of the
bosonic Hamiltonian (8) and the fermionic Hamiltonian (1) are equal
for zero, one and two bosons and quasiholes, respectively.
For zero bosons, the constant $\tilde{E}(N)=E_{g}(N,N)$ is the energy of the
filled lowest
Landau level. For one boson, we find the relation $\tilde{\varepsilon }(N,m)=
\varepsilon _{-}^{n}(N,M_{N+1}-m;1)$ with $0 \leq m \leq N-1$.
In the last step, we
determine the pseudopotential coefficients of the bosonic interaction
$\tilde{W}$
by requesting that the energies of (1) and (8) agree for the
two--quasihole state and the corresponding two--boson state.
Because the
energy eigenvalues of a two--boson system without
any boundary and with an interaction $\tilde{V}$ are (see above)
\begin{equation}
\tilde{E}_{i}(2,\tilde{M}) = \tilde{V}_{2i-2}
\end{equation}
with $ \tilde{M} \geq 0$ and $1 \leq i \leq [\frac{(\tilde{M}+2)}{2}]$ we
find by combining (8), (9) with (6), (7) in the leading order in $N$
\begin{equation}
N\varepsilon _{0}(1) + 2 \varepsilon _{-}^{n}(1) + \tilde{V}_{2i-2}
= V_{2i-1} - 4\varepsilon _{0}(1) + (N+2)\varepsilon _{0}(1) + O(1/\sqrt{N}).
\end{equation}
{}From (4) we know $\varepsilon _{0}(1) + \varepsilon _{-}^{n}(1)=0$
independent of the special choice of the fermionic interaction and thus
we get finally in the thermodynamic limit
\begin{equation}
\tilde{V}_{2i-2}=V_{2i-1}  \hspace{1.0cm} (i \geq 1).
\end{equation}
Doing this identification successively for $\tilde{M}=0,1,\ldots $ all
pseudopotential coefficients $\tilde{V}_{2i}$ are defined
uniquely. Particularly, for a finite number of non--zero pseudopotential
coefficients
$V_{2i-1}$, $\tilde{W}$ is restricted
too. From (11) it is obvious that if the short--range contribution in $W$ is
dominant this property
holds also for the quasihole interaction $\tilde{W}$.\\
It should be emphasized
that we do not perform a mapping relating fermionic and bosonic single particle
operators, but instead
a mapping from the fermionic energy eigenstates of the one-- and
two--quasihole
system to one-- and two--boson states in
the thermodynamic limit. This mapping could be constructed
because the dimensions of the Hilbert spaces are equal.\\

{\em Comparison of the ground state energies.} The hierarchical
theory assumes now that the Hamiltonian (8) with up to a two--particle
interaction correctly
describes the energetically low lying energy eigenstates,
even if a macroscopic number
of quasiparticles is present. Three--particle interactions and higher are
neglected.
According to Haldane \cite{Hald83}, starting from the
filled Landau level,
$\tilde{N}=N/p+1$ quasiholes (bosonic filling factor $1/p$) should form
again a stable
daughter state whose ground state energy per
electron $\tilde{\varepsilon _{0}}(\frac{1}{p+1})$ should then be equal to the
ground state energy per particle $\varepsilon _{0}(\frac{p}{p+1})$
at the filling factor
$\nu =\frac{p}{p+1}$ of the daughter state. In order to calculate
$\varepsilon _{0}(\frac{p}{p+1})$, we use the particle--hole--transformation,
see (5). Since for
our special choice (2) of the $V_{k}$ the ground state energy at $\nu
=\frac{1}{p+1}$ is zero, only the single particle term and the constant in
(5) contribute and the ground state energy per particle at
$\nu =\frac{p}{p+1}$ is
\begin{equation}
\varepsilon _{0}(\frac{p}{p+1})=\lim _{N \to \infty}\frac{E_{g}(N,N_{\Phi
}=\frac{(p+1)}{p}N+1)}{N}=\frac{(p-1)}{p}\varepsilon _{0}(1).
\end{equation}
This fermionic energy per particle has to be
compared with the ground state energy per number of electrons
$\tilde{\varepsilon} _{0}(\frac{1}{p})$ in the thermodynamic limit
resulting from the bosonic Hamiltonian (8) at
filling factor $1/p$. But $\tilde{V}_{2i}=0$
for all $i \geq p/2$, i.\
e.\ , a Laughlin wave function  with $q=p$ is the exact ground state wave
function, and the interaction term contributes zero to the energy.
The ground state energy per electron
is for this state with $\tilde{N}$ bosons (cf.\ of (8))
\begin{equation}
\tilde{\varepsilon }_{0}(\frac{1}{p})
=
\lim_{N \to \infty}\frac{\tilde{E}_{g}(\tilde{N}=
\frac{(N+p-1)}{p},N_{\Phi }=N)}{N} = \frac{(p-1)}{p} \varepsilon _{0}(1).
\end{equation}
Thus, the ground state energies per particle in the fermionic
and the hierarchical description, respectively, come out
exactly the same. \\

{\em Discussion.} In summary, we have presented an explicit analytical
calculation for the quasiholes at $\nu =1$ which
corraborates the hierarchical theory by comparing exactly ground state energies
per particle in
the fermionic and the effective bosonic model description. This was done by
an exact mapping of the fermionic two--quasihole Hilbert space onto a
two--boson Hilbert space. The matrix elements of the resulting
Hamiltonian (8) with up to a two--particle interaction were determined
analytically in
the thermodynamic limit. Crucial for
the analytical treatment of this case was the property $\varepsilon
_{0}(1) + \varepsilon _{-}^{n}(1)=0$ which led to the simple relation
(11) for the pseudopotential coefficients. The calculation showed
that at least for the ground state energy per particle, the description by an
effective Hamiltonian (8) containing only two--boson interactions is
successful. The question whether the lowest
excited states are also describable by this Hamiltonian can not be answered
within the framework of this analytical treatment. \\
There are at
least two interesting extensions of this work.
The first one should generalize our model
to a model with Coulomb interaction and
background. However, the ground state energy per particle at
filling factor $1/p$ (e.\ g.\ 1/2) of the Hamiltonian (8) in such a case
has to be determined
numerically by extrapolating the results of finite $N$ calculations because
the ground state energy of the interaction term is unknown. The second
extension concerns the quasiparticles at parent states $1/q$ ($q \neq
1$).
In this case, our
approach can serve as a starting point for generalizing the scheme
determining the pseudopotential coefficients from two--quasiparticle
spectra. Thus, this scheme should be applied to the quasiparticles
at $\nu =1/3$ which have already been treated by other methods
\cite{BM91,ET92}.
The pseudopotential
coefficients of the interaction have then to be extracted from finite $N$
spectra
and an additional extrapolation becomes necessary. Furthermore, the
mapping of the Hilbert space is then restricted to the lowest energy levels
in each of the blocks where two--quasiparticle components occur, while in the
present $\nu =1$--case all energy levels could be mapped. \\    \indent
I gratefully acknowledge stimulating discussions with W.\ Apel, as well as
his critical reading of the manuscript.

\end{document}